\documentclass[manuscript,screen,nonacm]{acmart}

\AtBeginDocument{%
  }

\setcopyright{none}
\renewcommand\footnotetextcopyrightpermission[1]{}
\begin{document}

\title{"Everyone Says Them": Deception Typologies, Probabilistic Trust, and Grassroots Safety Knowledge Among Gay Dating App Users in China}

\author{Yibo Meng*}
\affiliation{%
  \institution{Tsinghua University}
  \city{Beijing}
  \country{China}
}

\author{Lyumanshan Ye*}
\affiliation{%
  \institution{Shanghai Jiao Tong University}
  \city{Shanghai}
  \country{China}
}

\author{Yingfangzhong Sun*}
\affiliation{%
  \institution{Politecnico di Milano}
  \city{Milan}
  \country{Italy}
}

\author{Bingyi Liu}
\affiliation{%
  \institution{University of Michigan}
  \city{Ann Arbor}
  \country{United States}
}

\author{Huidi Lu}
\affiliation{%
  \institution{University of Oxford}
  \city{Oxford}
  \country{United Kingdom}
}

\author{Xiaolan Ding}
\affiliation{%
  \institution{North China University of Science and Technology}
  \city{Tangshan}
  \country{China}
}

\renewcommand{\shortauthors}{Meng et al.}

\thanks{*These authors contributed equally to this work.}

\begin{abstract}
Gay dating applications have become critical platforms for sexual minority men to seek relationships and community, yet they also expose users to deceptive interactions that remain underexplored in HCI and CSCW research. This study examines how gay male users in China experience, identify, and respond to deception on dating applications. Through semi-structured interviews with 22 participants across platforms including Blued, Aloha, Fanka, and Soul, we make three contributions. First, we identify a typology of deceptive practices extending beyond profile misrepresentation to encompass relational, emotional, financial, and commercial forms of deception. Second, we document the layered, probabilistic verification strategies users develop through long-term platform use, showing that trust assessment operates as a multi-signal, provisional process rather than a binary judgment. Third, we demonstrate that risk recognition is a collaborative practice shaped by the circulation of experience, the abstraction of recurrent tactics, and the codification of shared rules within the community.
\end{abstract}

\keywords{deception, gay dating apps, trust, community knowledge, China, LGBTQ+, online safety, qualitative study}

\maketitle

\section{Introduction}

Dating applications hold particular significance for gay men in 
societies where homosexuality remains stigmatized: platforms such 
as Grindr, Blued, and Aloha provide relatively safe spaces for 
identity expression, partner seeking, and community 
building~\cite{blackwell2015, wu2020, wu2021, meng2026focus}. In China, where 
homosexuality was decriminalized in 1997 and removed from the 
classification of mental disorders in 2001 but social acceptance 
remains limited~\cite{kong2011, wei2007, zhao2025immersive}, gay dating applications 
serve as critical infrastructure for a population with few 
alternative channels for meeting potential partners.

The same affordances that make dating platforms valuable, including 
low-cost communication, selective self-presentation, and access to 
a large pool of strangers, also create fertile ground for 
deception~\cite{hancock2007, toma2008, walther2007, zhang2026pervasive}. For gay men 
in China, this ecology of deception is further shaped by 
distinctive structural conditions. The semi-closeted status of 
many users amplifies vulnerability to exploitation, and apps such 
as Blued and Aloha coexist with more general social platforms, 
creating a complex ecosystem in which users navigate different 
norms, audiences, and risks~\cite{wu2020, wu2023, luo2025s}. The forms of 
deception users encounter extend well beyond the profile-level 
misrepresentation that has dominated the literature: relational 
misrepresentation oriented toward sexual access, sustained 
emotional manipulation, financially motivated fraud, and commercial 
marketing disguised as ordinary interaction constitute a range that existing frameworks fail to capture.

Existing research has focused predominantly on heterosexual dating 
contexts in Western societies, and has paid relatively little 
attention to how users actively respond to deception. Studies of 
online dating have documented self-presentation 
strategies~\cite{ellison2006, chen2025gestobrush} and uncertainty reduction 
behaviors~\cite{corriero2016, gibbs2011, chen2026not}, but have less often 
examined the practical verification and coping strategies that 
users develop through accumulated platform experience. The role of 
community in shaping individual risk recognition has been largely 
overlooked. This study addresses these gaps through semi-structured 
interviews with 22 gay male users in China, asking:

\begin{itemize}
\item \textbf{RQ1.} What different kinds of deceptive interactions 
do users encounter on gay dating applications?
\item \textbf{RQ2.} How do users assess credibility and develop 
verification and coping strategies?
\item \textbf{RQ3.} How are experiences of deception circulated 
and transformed into shared judgment knowledge within the community?
\end{itemize}

\section{Related Work}

Research on online dating deception has established that 
misrepresentation is common but typically confined to profile-level 
discrepancies in height, weight, and 
age~\cite{hancock2007, toma2008, cao2026causalinfluencemaximizationsteadystate}. Subsequent work has examined 
linguistic markers of deception~\cite{toma2012, cao2026beyond}, profile 
veracity~\cite{ellison2012, ma2026can}, and how platform affordances shape 
self-presentation~\cite{ranzini2017, walther1996, walther2007, zhang2025slideaudit}. 
Theoretical frameworks including Goffman's dramaturgical 
model~\cite{goffman1959, meng2026decoration} and Walther's hyperpersonal 
model~\cite{walther1996, walther2007, liu2025supporting} emphasize that online 
environments afford greater control over self-presentation, enabling 
both strategic identity construction and deliberate 
misrepresentation. Romance scam research has extended this picture 
by showing how sustained communication builds trust before financial 
requests are 
introduced~\cite{whitty2015, whitty2012, whitty2016, meng2026engagement}. This 
literature has, however, focused predominantly on heterosexual, 
Western contexts, leaving the ecology of deception on platforms 
used by sexual minority men largely unexamined.

HCI and CSCW research on LGBTQ+ communities has addressed identity 
management~\cite{devito2018, haimson2018, meng2026creating}, technology-mediated 
safety~\cite{scheuerman2018, su2026capnav}, and platform 
governance~\cite{taylor2024a, taylor2024b, su2025flymethrough}. Work in the Chinese 
context has traced relationship development on gay dating 
apps~\cite{wu2020, meng2026tibetcpr}, generational adoption of 
Blued~\cite{wu2021}, and the intersection of sex, risk, and 
stigma~\cite{wu2023}. Trust and uncertainty reduction in mobile 
dating have also been 
examined~\cite{corriero2016, gibbs2011, hallam2019}, with 
warranting theory offering a framework for understanding how users 
seek external verification of self-presented 
claims~\cite{ellison2012}. Our study builds on these foundations 
by examining deception as a processual, interactional phenomenon, 
and by foregrounding the community-based mechanisms through which 
risk recognition is collectively generated.

\section{Method}

This study employed semi-structured interviews with 22 gay male 
users of dating applications in China, ranging in age from 19 to 
60. Participants were recruited through snowball sampling via 
personal networks; the only inclusion criterion was prior experience 
using gay dating or social applications. Because the research 
involved sensitive identities and private experiences, recruitment 
through familiar social ties helped reduce concerns and increase 
willingness to participate. Participants came from both urban and 
rural backgrounds with diverse occupations and educational levels, 
providing variation in how different users experienced platform-based 
interaction. Most used multiple platforms simultaneously, with Blued, 
Aloha, Fanka, and Soul most frequently mentioned. Length of use 
ranged from one to nine years, with most participants having over 
three years of experience, sufficient for reflecting on how their 
understanding of interaction risks had evolved over time.

Interviews were conducted via online voice or video calls, each 
lasting approximately 50 to 60 minutes, and were audio-recorded 
with participants' consent before transcription and anonymization. 
Interviews proceeded from less sensitive to more sensitive topics 
across four stages: platform use background, problematic interaction 
experiences, risk judgment and coping strategies, and community 
experience and learning. Within each stage, participants were asked 
to describe specific situations, identify what triggered suspicion, 
and reflect on how experiences had circulated within their 
communities. The study received ethics approval from an anonymized 
institution, and each participant received 50 RMB as compensation. 
Data were analyzed using thematic 
analysis~\cite{bowman2023, braun2006}. Transcripts were independently 
coded by two researchers, who assigned descriptive labels to segments 
covering problematic interactions, judgment cues, coping behaviors, 
and community experience. Codes were iteratively compared and refined 
through discussion until thematic saturation was reached.

\section{Findings}

Our findings are organized around three interconnected dimensions: 
the types of deceptive interaction users encounter, the verification 
and coping strategies they develop in response, and the 
community-based processes through which risk recognition capacities 
are collectively generated.

\subsection{Types of Deceptive Interaction}

Participants described five forms of deception that differ in 
underlying intent and interactional logic. The most widely reported 
was \textit{pian pao} (``deceiving someone into sex''), mentioned by 
all 22 participants. This involved presenting oneself as seeking a 
serious relationship while pursuing short-term sexual access. What 
made it deceptive was not casual sex itself, but the deliberate 
mobilization of relational scripts, including talk of commitment, 
shared futures, and emotional investment, to accelerate intimacy 
and lower the other person's guard. As P2 observed: ``If someone 
makes it clear from the beginning that it is just about sex, that 
is actually okay. The problem is that some people will say they 
want a serious emotional relationship... but in the end it is just 
to get you into bed more quickly.''

Fifteen participants also described a more sustained form of 
emotional deception, in which the other person maintained long-term 
ambiguity through daily contact and emotional responsiveness without 
ever committing to relationship progression or meeting offline. P6 
captured this pattern: ``He would not completely reject you, but he 
also would not let you give up. He just kept giving you a little bit 
of response so that you felt maybe there was still a possibility.''

A third form, mentioned by seven participants, involved financial 
fraud embedded within relational contexts. Ordinary conversation 
eventually gave way to narratives of sudden hardship, investment 
opportunities, or appeals to mutual support within what appeared to 
be a developing relationship. One participant described the gradual 
buildup: ``It was not the kind of thing where you can tell at a 
glance that someone is a scammer. He would first present himself as 
very sincere, as if he were genuinely interested in you. Only after 
you started to feel that this person seemed okay would he slowly 
move the topic toward money'' (P14). Fraud narratives typically 
relied on one of three strategies: a short-term difficulty framed 
as requiring temporary assistance, an investment opportunity backed 
by apparent professional credentials, or a mutual support framing 
that positioned financial help as natural within what was presented 
as an already intimate bond.

A fourth form concerned accounts that entered the platform as 
ordinary users before redirecting conversation toward external 
platforms, paid content, or commercial services, which participants 
described as ``hanging up a sheep's head while selling dog meat.'' 
P14 articulated what participants found most objectionable: ``The 
most annoying thing is not that he is doing marketing. It is that 
he is clearly not here to meet people, but pretends as if he is 
seriously chatting with you.'' Finally, all participants identified 
false identity construction as a cross-cutting mechanism underlying 
the other forms: fabricated photos, inflated social status, and 
misrepresented biographical details served to establish 
attractiveness and trust before other forms of deception could take 
hold. Deception regarding sexual role was also noted. Role labels 
such as 0, 1, and 0.5 are used in the Blued community to 
indicate preferred roles in intimate relationships. P22 
recalled: ``He wrote on his profile that he was 0.5 or 1, 
but when we actually met he was completely 0.''
\subsection{Verification and Coping Strategies}

Faced with persistent uncertainty about others' intentions and 
identity, participants developed a layered set of verification 
practices that unfolded gradually across the course of interaction. 
Rather than seeking definitive proof, they engaged in what can be 
characterized as probabilistic inference, continuously integrating 
multiple signals into an evolving judgment about whether to proceed.

When evaluating relational intent, 15 participants described 
heightened wariness toward emotional escalation that outpaced mutual 
understanding. Phrases such as ``you are different from other 
people'' early in an interaction were treated not as romance but as 
signals requiring further observation. Participants also monitored 
whether verbal commitments translated into real-world progression, 
and some deliberately slowed the pace of interaction to test the 
other person's sustained interest. P21 noted: ``Later I would 
deliberately slow things down a bit, like not meeting so quickly. 
If the other person cools off right away, that actually says a lot.''

On financial risk, many participants had developed firm personal 
rules, typically after encountering fraud directly or hearing about 
it from others. P16 stated: ``As long as we haven't met many times, 
once money enters the conversation, it's basically over.'' The 
underlying principle was that platform-based familiarity does not 
constitute real-world relational obligation. Identifying commercial 
promotion accounts was generally described as the most 
straightforward task: such accounts quickly redirected conversation 
toward external platforms, showed little interest in reciprocal 
exchange, and followed recognizable patterns that participants had 
learned to detect and dismiss.

Identity verification followed a graduated sequence: asking 
follow-up questions to probe the coherence of biographical claims, 
requesting multiple photos in different settings to assess 
consistency, and, when a meeting seemed plausible, asking for a 
real-time photo or initiating a video call. Video calls were 
regarded as the most reliable form of confirmation, simultaneously 
verifying appearance, real-time presence, and interactional 
continuity. Participants noted, however, that proposing a video 
call too early could signal distrust and disrupt the relational 
atmosphere. As P13 explained: ``Video is the most direct way, but 
it is also not something you can bring up right away. Usually it is 
after we have chatted for a while and I feel like we might meet 
that I will ask whether we should video first.'' For attributes 
such as educational background and income level, participants 
acknowledged that no reliable verification method existed, and 
that sustained conversational observation remained the primary 
resource: ``If it is fake, sooner or later it will show. If you 
keep talking long enough, you will eventually notice places where 
the logic breaks down'' (P4).

\subsection{Community-Based Risk Recognition}

Nineteen participants stated that their judgment capacities had not 
developed through individual experience alone. Formal platform 
mechanisms such as reporting systems and account verification were 
regarded as inadequate for the ambiguous, gray-area risks that 
dominated everyday use. Community exchange filled this gap.

When participants encountered suspicious interactions, the first 
response was typically not to report to the platform but to 
describe the experience to friends or share screenshots in WeChat 
groups, seeking both interpretation and validation. P19 noted: 
``Sometimes when you are inside it yourself, you cannot see clearly. 
But a friend looks at it and says this really looks like one of 
those \textit{taolu} (routine patterns), and I suddenly realize 
what is going on.'' Through repeated narration and comparison, 
individual cases were abstracted into recognizable 
patterns, \textit{taolu}, that described not specific bad actors 
but recurring interactional logics: rapid intimacy building 
followed by a push toward meeting as a common path of 
\textit{pian pao}; sustained companionship with consistent 
avoidance of commitment as a typical rhythm of emotional deception. 
These patterns eventually crystallized into shared rules that 
circulated informally but were widely observed: do not transfer 
money to someone not yet met in person; request a video call 
before agreeing to meet; treat rapid romantic escalation with 
caution. P3 described the character of such rules: ``A lot of 
things are not taught to you explicitly by one person. It is just 
that everyone says them, and after hearing them enough, you realize 
they are default rules.''

Risk learning in the community operates as an ongoing cycle rather 
than a one-time input. New users enter the knowledge system through 
others' accumulated experience; as they gather experience of their 
own, they contribute in turn to its circulation and revision. Many 
participants reflected that they had not initially possessed these 
judgment resources and that long-term community immersion had been 
essential to developing them. P14 recalled: ``When I first started 
using these apps, I was actually quite simple-minded. I would 
easily believe what others said. Later, after hearing about what 
my friends had encountered and almost stepping into things myself, 
I slowly learned how to judge.'' Participants were equally willing 
to pass this knowledge on to newer users: ``In the past, other 
people reminded me. Now, if I have friends who are just starting 
to use these apps, I will also tell them some basic things, like 
don't trust too quickly, don't transfer money easily'' (P2). Risk 
recognition is thus not merely acquired but actively reproduced 
through community practice, transforming what begins as individual 
vulnerability into a collectively maintained capacity for 
navigating deception.

\section{Discussion and Conclusion}

\subsection{Trust as Ongoing Uncertainty Management}

Prior research has largely framed trust in online dating as a 
matter of verifying the accuracy of others' self-presentation~\cite{ellison2012, hancock2007}. 
Our findings complicate this picture. In the context of gay dating 
applications in China, definitive verification is rarely achievable: 
participants described continuously assessing, revising, and 
updating their judgments as interactions unfolded. Trust did not 
emerge from the confirmation of objective facts but from a situated 
and provisional evaluation of risk, a level of confidence deemed 
sufficient to proceed to the next step. This reframes trust from a 
binary outcome into an ongoing process of uncertainty 
management~\cite{corriero2016, gibbs2011}, shaped by the temporal 
progression of interaction rather than any single decisive piece 
of evidence. The graduated verification strategies participants 
developed, from conversational probing to photo consistency checks 
to real-time video confirmation, reflect not a pursuit of certainty 
but a practical navigation of irreducible uncertainty.

\subsection{Deception as Intentional Misalignment}

Existing literature has focused predominantly on discrepancies 
between presented information and objective reality: inaccurate 
profiles, fabricated biographical 
details~\cite{ellison2006, toma2010, toma2012}. Our data suggest 
a different and broader conception is needed. The forms of 
deception participants found most difficult to name and most 
damaging to experience, including \textit{pian pao} and sustained 
emotional ambiguity, involved no explicitly false statements. The 
deception lay instead in the misalignment between expressed and 
actual relational intent. Presenting oneself as seeking a serious 
relationship while pursuing short-term sexual access, or 
maintaining prolonged emotional engagement with no intention of 
developing the relationship, is experienced as deeply deceptive 
even when no verifiable lie has been told. This shifts analytical 
attention from factual accuracy to relational 
alignment~\cite{goffman1959}: what users must navigate is not only 
what others claim about themselves but what those others actually 
intend within the interaction.

\subsection{Community Knowledge as Compensation for Platform 
Limitations}

Platform safety mechanisms typically assume that risk is 
identifiable, reportable, and rule-bound. The risks our 
participants described rarely conform to this model: \textit{pian 
pao} violates no written policy; emotional manipulation leaves no 
reportable evidence; a marketing account engaging in ordinary 
conversation is, on the surface, indistinguishable from any other 
user. Formal governance structures are thus inadequate 
for the gray-area risks that dominate everyday 
use~\cite{scheuerman2018}. Users compensate by constructing an 
informal, distributed system of risk recognition outside the 
platform: sharing screenshots, narrating encounters, abstracting 
recurring patterns into \textit{taolu}, and codifying shared rules 
that circulate through friendship networks and community 
groups~\cite{devito2018, haimson2018}. This community-based 
knowledge production is not incidental to platform safety; it is 
doing much of the safety work that platforms cannot.

These findings carry design implications. Current reporting 
mechanisms ask users to identify a specific violation, a 
requirement that structurally filters out the ambiguous, processual 
risks that dominate everyday use. Platforms could instead support 
users' existing sensemaking practices by surfacing contextual cues 
that users already attend to but cannot easily access: how long an 
account has been active, how quickly intimacy language has appeared 
relative to the duration of contact, or how communication patterns 
have shifted over time. A mechanism that allows users to flag an 
interaction as uncertain or uncomfortable, without requiring them 
to prove wrongdoing, could better capture the texture of everyday 
risk. Structured spaces for anonymized experience sharing could 
extend the reach of community knowledge beyond existing social 
networks, though any such feature would need to preserve the 
organic, peer-driven character that makes community knowledge 
effective in the first place.

This study demonstrates that deception on gay dating platforms in 
China is not primarily a matter of false information but of 
intentional misalignment, that trust assessment operates as 
layered and probabilistic uncertainty management rather than binary 
verification, and that the capacity to navigate deception is a 
socially generated competence produced through community 
circulation and collective interpretation. For platforms serving 
marginalized users whose access to alternative channels is 
structurally constrained, recognizing and supporting these 
grassroots practices represents a more adequate foundation for 
safety than rule-based enforcement alone.

\section*{Generative AI Use Disclosure}
This work did not use generative AI tools in the research, analysis, or writing process.

\bibliographystyle{ACM-Reference-Format}

\bibliography{references}

\end{document}